\documentclass[twocolumn,aps,pra,showpacs,superscriptaddress]{revtex4-1}

\usepackage{epsfig}
\usepackage{dcolumn}
\usepackage{amssymb}
\usepackage{color}
\usepackage{amsmath}

\begin{document}

\title{Bose-Einstein condensation of photons with nonlocal nonlinearity in a dye-doped graded-index microcavity}

\author{Marcello Calvanese Strinati}
\email{marcello.calvanesestrinati@gmail.com}
\affiliation{Dipartimento di Fisica, Universit\`{a} di Roma ``La Sapienza'', Piazzale Aldo Moro 5, 00185 Rome, Italy}

\author{Claudio Conti}
\email{claudio.conti@uniroma1.it}
\homepage{http://www.complexlight.org/}
\affiliation{Dipartimento di Fisica, Universit\`{a} di Roma ``La Sapienza'', Piazzale Aldo Moro 5, 00185 Rome, Italy}
\affiliation{Institute for Complex Systems (ISC-CNR), National Research Council,  Via dei Tarini 19, 00185 Rome, Italy}

\begin{abstract}
We consider a microcavity made by a graded-index (GRIN) glass, doped by dye molecules, placed within two planar mirrors and study Bose-Einstein condensation (BEC) of photons. The presence of the mirrors leads to an effective photon mass, and the index grading provides an effective trapping frequency; the photon gas becomes formally equivalent to a two dimensional Bose gas trapped in an isotropic harmonic potential. The inclusion of nonlinear effects provides an effective interaction between photons. We discuss, in particular, thermal lensing effects and nonlocal nonlinearity, and quantitatively compare our results with the reported experimental data.
\end{abstract}
\date{\today}

\maketitle

\section{Introduction}
An experiment reported in 2010 \cite{natureKlaers:2010,naturephysicsKlaers:2010,klaersweitz} demonstrated BEC of photons, and  was realized by using a dye-filled optical microcavity. The microcavity affects the photon energy-momentum relation with respect to free space and introduces an effective photon mass, thus creating the conditions for the BEC.

Jan Klaers \emph{et. al.} \cite{klaersweitz} showed this property in a system made up of two spherical mirrors placed at a distance $D_0$ measured on the optical axis. If $R$ is the radius of curvature of the mirrors and $r=|\mathbf{r}|=\sqrt{x^2+y^2}$ is the distance from the optical axis (on the $xy$ plane) one can verify that the energy-momentum relation in the paraxial approximation reads as
\begin{equation}
E(\mathbf{k},\mathbf{r})\simeq\frac{mc^2}{n_0^2}+\frac{\hbar^2\mathbf{k}_r^2}{2m}+\frac{1}{2}m\Omega^2\mathbf{r}^2-\frac{mc^2}{n_0^3}n_2I(\mathbf{r}) \label{photonhamiltonian}\,\, ,
\end{equation}
where  $\mathbf{k}_r$ is the transverse wavenumber, $n_0$ is the index of refraction, $m$ is the effective photon mass, $c$ is the speed of light in free space, $\Omega$ is the effective trapping frequency, and the last term arises from the optical Kerr effect. The effective photons mass $m$ and the effective trapping frequency $\Omega$ are
\begin{equation}
m=\frac{\hbar\pi qn_0}{cD_0} \qquad \Omega=\frac{c}{n_0}\sqrt{\frac{2}{D_0R}} \label{effectivemf}\,\, .
\end{equation}
In \cite{klaersweitz} the distance between the two mirrors was $D_0\simeq1.46\,\mathrm{\mu m}$, and the corresponding free spectral range was of order $10^{14}\,\mathrm{Hz}$ and comparable with the spectral linewidth of the dye. Only one longitudinal mode with order $q=7$ was within the dye spectral linewidth, actually freezing out a degree of freedom \cite{chiao1999,chiao2000}. In this configuration the photons gas can be treated as a two dimensional Bose gas trapped in an isotropic harmonic potential, being the longitudinal mode number fixed, and neglecting the nonlinear term, Eq. (\ref{photonhamiltonian}) may be interpreted as the Hamiltonian of a two dimensional harmonic oscillator with energy spectrum
\begin{equation}
\epsilon_{rs}=\frac{mc^2}{n_0^2}+\hbar\Omega\left(r+s+1\right) \label{harmonicoscillatorenergies}\,\, ,
\end{equation}
where $r$ and $s$ are two non-negative integer numbers. For such a system, BEC is expected to occur following the previous investigations \cite{RevModPhys.71.463,Pitaevskii.Stringari:BoseEinsteinConsensation,pethick2008bose,refId0,2014arXiv1402.0706Y} in ultra-cold trapped atoms. The realization of the experiment has opened new perspectives in the field of photonics, including novel theoretical investigations \cite{PhysRevA.88.013615,PhysRevE.88.022132,2014arXiv1405.7030C,PhysRevA.89.033862,connaughton:263901} as well as new experimental proposals \cite{2014arXiv1406.1403D}.

A signature of the photon BEC is the appearance of the condensate peak (photons massively occupying the ground $\mathrm{TEM}_{00}$ mode) over the much broader underlying thermal component (photons emitted by fluorescence over all the other possible modes sustained by the cavity). This phase transition was found to occur at the critical number of photons expected for a BEC.

In \cite{klaersweitz} by measuring the condensate diameter as a function of the number of photons on the ground mode (condensate fraction) a \emph{broadening} of the condensate diameter was observed and ascribed to a \emph{repulsive} interaction between photons. This was explained by the nonlinearity arising from a thermo-optical effect, such as thermal lensing, and modelled by the Gross-Pitaevskii equation (GPE) with the assumption of a \emph{contact}, or \emph{local}, interaction.

\begin{figure}[t]
\centering
\includegraphics[width=6.9cm]{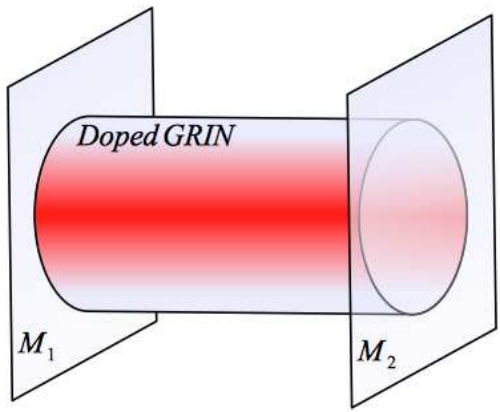}
\caption{(Color online) Scheme of the graded-index (GRIN) microcavity, a dye-doped GRIN lens is placed between two planar mirrors ($M_1$ and $M_2$).}
\label{fig:grincavity}
\end{figure}

In this work, we theoretically and numerically study the BEC of photons including a finite degree of nonlocality. Instead of considering a bispherical microcavity we make reference to a graded index medium sketched in Fig. \ref{fig:grincavity}. This configuration is equivalent to considering the case of \cite{klaersweitz}, but has the advantage of a more direct derivation of the BEC formalism and indicates an alternative experimental configuration.

Another interesting feature for the GRIN case is the fact that the thermal nonlocality can be either focusing or defocusing, thus opening the possibility for a variety of fundamental studies. The nonlocal response of the medium is actually a natural effect when thermal or diffusive type of nonlinearity are considered \cite{PhysRevA.48.4583} in optical systems. In the case of BEC in ultra-cold trapped atoms the local form of the interaction potential is considered in many cases of interest \cite{hadzibabic:11}, however nonlocality between atoms has to be included when a finite-range interaction is taken into account \cite{PhysRevA.57.R3180,PhysRevA.61.051601,PhysRevE.62.4300}.

It is interesting to observe that nonlocal effect for BEC of photons may be more relevant with respect to standard ultra-cold atomic clouds because of the different origin of the nonlinearity.

For the sake of completeness we first review below the local case, and validate the nonlocal generalization by comparison with experimental data. The analysis is made with reference to a GRIN microcavity which we propose as an alternative setup for the BEC of photons, but also holds for the case in Ref. \cite{klaersweitz}.

This manuscript is organised as follows: in Section \ref{sec:photonsincavity} we discuss the photon wave function in the absence of nonlinearity, and in Section \ref{sec:nonlineareffects} we introduce the effects of nonlinearity. In Section \ref{sec:localnonlinearity}, for the sake of completeness, we review the analysis with local nonlinarity. In Section \ref{sec:nonlocalnonliearity} we discuss the nonlocal nonlinearity and compare with the reported experimental results. Conclusions are drawn in Section \ref{sec:conclusions}.

\section{Photons in a GRIN cavity}
\label{sec:photonsincavity}
We consider a microcavity made by a quadratic-index glass doped with dye molecules placed between two planar mirrors separated by a distance $D$. Let $z$ be the longitudinal direction; 
the index of refraction is written as
\begin{equation}
n(\mathbf{r})=n_0-\frac{1}{2}n_{2,L}\mathbf{r}^2 \,\, ,
\end{equation}
where $n_{2,L}>0$. We assume the index of refraction to vary only in the transverse ($x,y$) directions, we have 
\begin{equation}
n^2(\mathbf{r})\simeq n^2_0-n_0n_{2,L}\left(x^2+y^2\right) \label{grincavityindexofrefraction}\,\, .
\end{equation}
The Helmholtz equation in a GRIN medium \cite{saleh2013fundamentals} can be written as
\begin{equation}
\left[\nabla^2+k^2_0n^2(x,y)\right]E(x,y,z)=0 \label{helmholtzequationgrinmedium}\,\, ,
\end{equation}
where $k_0=2\pi/\lambda$ is the wave vector in free space; one can assume the field propagating along the $z$ axis so that its dependence on $z$ is given by a factor $\exp(i\beta z)$, where $\beta$ is the propagation constant. Equation~(\ref{helmholtzequationgrinmedium}) becomes
\begin{eqnarray}
&&\left[\frac{\partial^2}{\partial x^2}+\frac{\partial^2}{\partial y^2}-k^2_0n_0n_{2,L}\left(x^2+y^2\right)\right]E(x,y) \nonumber\\
\nonumber\\
&&=\left(\beta^2-k^2_0n^2_0\right)E(x,y) \label{transversegrinequation} \,\, ,
\end{eqnarray}
and by introducing the following quantities
\begin{equation}
\xi={\left(n_0n_{2,L}k^2_0\right)}^{1/4} \quad u=\xi x \quad v=\xi y \quad \Lambda=\frac{k^2_0n^2_0-\beta^2}{2\xi^2} \label{helmholtzgrinvalues}\,\, ,
\end{equation}
equation (\ref{transversegrinequation}) can be recast in the form
\begin{equation}
\left[-\frac{1}{2}\left(\frac{\partial^2}{\partial u^2}+\frac{\partial^2}{\partial v^2}\right)+\frac{1}{2}\left(u^2+v^2\right)\right]E(u,v)=\Lambda E(u,v) \label{hogrinequation}\,\, ,
\end{equation}
which is formally the Schr\"{o}dinger equation for a two-dimensional harmonic oscillator. This scheme is well known to yield the Hermite-Gauss modes, and the discrete set of values of the propagation constant is found to be in the paraxial approximation
\begin{equation}
\beta_{rs}\simeq k_0n_0-\sqrt{\frac{n_{2,L}}{n_0}}(r+s+1) \label{betagrinvalue} \,\, ,
\end{equation}
where $r$ and $s$ are two non-negative integer numbers. Being the GRIN medium placed within a planar-mirror resonator there is another condition that has to be considered: in a round-trip, the change of phase must be $2\pi q$, where $q$ is an integer number, and then this forces $\beta$ to be in turn discretized as ($q=1,2,...$)
\begin{equation}
\beta_q=\frac{\pi q}{D} \label{longitudinalmodediscretizationb}\,\, .
\end{equation}
By writing $k_0=\epsilon/\hbar c$ one can obtain the energy spectrum of the photon inside the GRIN cavity
\begin{equation}
\epsilon_{qrs}=\frac{\hbar c\pi q}{n_0D}+\frac{\hbar c}{n_0}\sqrt{\frac{n_{2,L}}{n_0}}(r+s+1) \label{energygrinspectrum}\,\, ,
\end{equation}
which can be rewritten by introducing an effective photon mass $m$ and an effective trapping frequency $\Omega$
\begin{equation}
m=\frac{\hbar\pi qn_0}{cD} \qquad \Omega=\frac{c}{n_0}\sqrt{\frac{n_{2,L}}{n_0}} \label{effectivegrinmassandfrequency} \,\, ,
\end{equation}
and then equation (\ref{energygrinspectrum}) becomes
\begin{equation}
\epsilon_{qrs}=\frac{mc^2}{n^2_0}+\hbar\Omega(r+s+1) \label{enrgygrinmassandfrequency}\,\, .
\end{equation}
Therefore, the photons gas becomes formally equivalent to a two dimensional Bose gas trapped in an isotropic harmonic potential, for which BEC is expected to occur. The energy spectrum given by (\ref{enrgygrinmassandfrequency}) is as in Eq.~(\ref{harmonicoscillatorenergies}) with the same definition of the mass and with a trapping frequency that in this case depends $n_{2,L}$ rather than on the curvature of the mirrors. As usual, the longitudinal mode number can be fixed by selecting a single longitudinal mode, hence the mass is a constant quantity and the energy spectrum (\ref{enrgygrinmassandfrequency}) becomes independent of $q$.

The characteristic length scale of the modes is $\xi^{-1}$. For the ground mode ($r,s=0$) this corresponds to the characteristic length scale of the harmonic oscillator $\sqrt{\hbar/m\Omega}$, as can be verified by using the definition of the effective photon mass $mc^2/n_0^2=h\nu_0$, where $\nu_0$ is the cutoff frequency, and by using Eq. (\ref{effectivegrinmassandfrequency}). The condensate diameter measured with the full width half maximum (FWHM) is $d=2\sqrt{\hbar\log(2)/m\Omega}$. A possible way to estimate the value of $n_{2,L}$ needed to fit the Klaers' \emph{et. al.} data is by requiring the condensate spot diameter
\begin{equation}
d={\left(\frac{2\log(2)\lambda}{\pi\sqrt{n_0n_{2,L}}}\right)}^{1/2}
\end{equation}
is the same measured in the experiment, being $\lambda$ the cut-off wavelength, which gives $n_{2,L}\simeq1.3\times10^6\,\mathrm{m^{-2}}$, where $n_0=1.33$, $\nu_0=506\,\mathrm{THz}$ and $d=14\,\mathrm{\mu m}$ were used; this value gives a trapping frequency $\Omega/2\pi\simeq3.5\times10^{10}\,\mathrm{Hz}$. The GRIN medium could be an alternative setup for Bose-Einstein condensation of photons; here, in particular, we considered the behaviour of the photons gas without including an effective interaction between photons. The role of nonlinearity is discussed in detail in the following sections.

\section{Nonlinear effects in BEC of photons}
\label{sec:nonlineareffects}
In the experiments in \cite{klaersweitz} the broadening of the condensate diameter was ascribed to the nonlinearity due to thermal effects inside the cavity. The nonlinear refractive index perturbation, treated as a self-interaction term between photons, was written as $\Delta n=n_2I(\mathbf{r})$, in a local form such that the variation of the refractive index in a point $\mathbf{r}$ is determined by the optical intensity in the same point.
\begin{figure}[t]
\centering
\includegraphics[width=8cm]{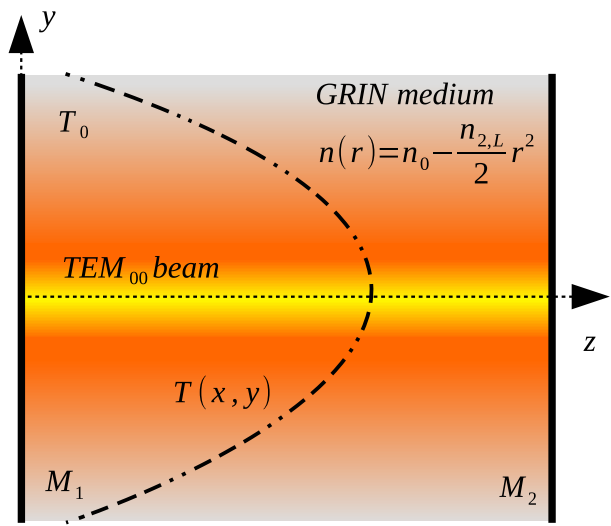}
\caption{(Color online) Scheme of the GRIN microcavity in presence of thermal effects: the massively occupied ground mode field ($\mathrm{TEM_{00}}$) heats up the medium in the vicinity of the optical axis causing a non-zero temperature gradient between the centre of the cavity and the outer space.}
\end{figure}

The thermo-optical effect can be discussed as follows \cite{Rusconi:04,PhysRevLett.95.213904,IturbeCastillo:96}: the electric field inside the cavity has a spatial extent  $W$, and the intensity heats up the medium inside the cavity causing a non zero temperature variation $\Delta T$ between the center of the optical cavity, where the temperature reaches its maximum, and the outer space, where temperature is kept constant to a value $T_0$, e.g. room temperature. The temperature gradient causes a change of the index of refraction which, for small $\Delta T$, can be written as
\begin{equation}
\Delta n_0=\frac{\partial n_0}{\partial T}\Delta T \,\, .
\end{equation}
The beam inside the cavity behaves like a heat source, and in the stationary regime $\Delta T$ obeys the transport equation
\begin{equation}
-\kappa\nabla^2\left(\Delta T\right)=\alpha I(\mathbf{r})=\frac{\alpha n_0}{2Z_0}{\left|E(\mathbf{r})\right|}^2 \label{transportequation}\,\, ,
\end{equation}
where $I(\mathbf{r})$ is the beam intensity, $\kappa$ is the thermal conductivity, $Z_0$ is the impedance of free space and $\alpha$ is the absorption coefficient. The solution to Eq. (\ref{transportequation}) is
\begin{equation}
\Delta T(\mathbf{r})=\int d\mathbf{r}'\,G\left(\mathbf{r}-\mathbf{r}'\,\right){\left|E\left(\mathbf{r}'\,\right)\right|}^2 \label{grinnonlocalprofile}\,\, ,
\end{equation}
where $G\left(\mathbf{r}-\mathbf{r}'\,\right)$ is the Green's function, and its form depends on the geometry of the specific device and on the boundary conditions. Equation (\ref{grinnonlocalprofile}) in general furnishes the nonlocal dependence of the temperature gradient on the field $E(\mathbf{r})$.

The variation of the index of refraction is then
\begin{equation}
\Delta n_0=\frac{\partial n_0}{\partial T}\int d\mathbf{r}'\,G\left(\mathbf{r}-\mathbf{r}'\,\right){\left|E\left(\mathbf{r}'\,\right)\right|}^2 \,\, .
\end{equation}
The variation of the index of refraction is always negative for gaseous and liquid materials \cite{Jurgensen:95}, whilst it can either negative or positive for solids, depending on their properties
(see, e.g., \cite{Gentilini:14}). This suggests that the index of refraction in the GRIN medium can be then written in the form
\begin{equation}
n(\mathbf{r})=n_0-\frac{1}{2}n_{2,L}\mathbf{r}^2+\int d\mathbf{r}'\,K\left(\mathbf{r}-\mathbf{r}'\,\right){\left|E\left(\mathbf{r}'\,\right)\right|}^2 \label{nonlocalgrinindex}\,\, ,
\end{equation}
where the integral kernel $K(\mathbf{r})$ in (\ref{nonlocalgrinindex}) is related to the Green function $G(\mathbf{r})$ in (\ref{grinnonlocalprofile}). The presence of a nonlocal nonlinearity in the index of refraction leads to the equation for the transverse electric field, which is formally the nonlocal Gross-Pitaevskii equation with an harmonic external potential. A similar study, however in a different context, was developed in \cite{PhysRevE.66.046619,PhysRevE.63.016610,turitsyn:85} regarding solitons in nonlocal nonlinear media, and in \cite{litvak:75} for thermal nonlocal self effects of wave beams in plasma. The nonlocal term in (\ref{nonlocalgrinindex}) plays the role of the interaction term beyond the typically adopted contact interaction approximation. By assuming a specific form of the Green function in (\ref{grinnonlocalprofile}) and by taking the electric field distribution as the ground mode one $I(\mathbf{r})=I_0\exp(-\mathbf{r}^2/W^2)$, the integral (\ref{grinnonlocalprofile}) can be carried out directly, leading to an explicit expression for the temperature gradient.
A possible model is the one proposed by Gordon \emph{et. al} \cite{:/content/aip/journal/jap/36/1/10.1063/1.1713919,Rusconi:04} where the temperature gradient (\ref{grinnonlocalprofile}) was found by assuming a infinite cylindrically symmetric medium with no convection.

\section{Local nonlinearity approximation}
\label{sec:localnonlinearity}
We start considering a \emph{local} response of the medium in Eq. (\ref{nonlocalgrinindex})
\begin{equation}
K\left(\mathbf{r}-\mathbf{r}'\,\right)=\frac{1}{2}n_{2,NL}\,\delta\left(\mathbf{r}-\mathbf{r}'\,\right) \,\, ,
\end{equation}
which gives the expression for the index of refraction
\begin{equation}
n(\mathbf{r})=n_0-\frac{1}{2}n_2\left(x^2+y^2\right)+\frac{1}{2}n_{2,NL}{\left|E(\mathbf{r})\right|}^2 \,\, .
\end{equation}
The wave equation in the GRIN medium with the local nonlinearity then becomes
\begin{eqnarray}
&&\left[\nabla^2_\perp-k^2_0n_0n_{2,L}\mathbf{r}^2+k^2_0n_0n_{2,NL}{|E(x,y)|}^2\right]E(x,y)\nonumber\\
\nonumber\\
&&=\left(\beta^2-n^2_0k^2_0\right)E(x,y) \label{nonlinearhelmholtzgrinequation}\,\, ,
\end{eqnarray}
where $\displaystyle{\nabla^2_\perp=\sum_{i=1}^{2}\frac{\partial^2}{\partial x_i^2}}$ and $\mathbf{r}$ indicates the two dimensional spatial position. As above, the electric field in (\ref{nonlinearhelmholtzgrinequation}) was assumed to be $E(\mathbf{r})=E(x,y)\exp(i\beta z)$. Here the electric field is normalized such that
\begin{equation}
\frac{\varepsilon}{2}\int d\mathbf{r}\,{\left|E(\mathbf{r})\right|}^2=\mathcal{E} \label{electricfieldnormalization}\,\, ,
\end{equation}
being $\varepsilon$ the dielectric constant and $\mathcal{E}$ the total energy on the ground mode; the total energy per unit length is related to the intensity distribution with
\begin{equation}
\frac{\mathcal{E}}{D}=\frac{n_0}{c}\int dxdy\,I(x,y) \,\, .
\end{equation}
By means of (\ref{helmholtzgrinvalues}) and by defining a dimensionless electric field and a dimensionless coupling constant
\begin{equation}
\tilde{E}(u,v)=\sqrt{\frac{\varepsilon D}{2\xi^2\mathcal{E}}}\,E(u,v) \qquad \tilde{g}=-\frac{k^2_0n_0n_{2,NL}\mathcal{E}}{\varepsilon D} \label{nonlineargrinvalue}\,\, ,
\end{equation}
equation (\ref{nonlinearhelmholtzgrinequation}) can be rewritten as
\begin{eqnarray}
&&\left[-\frac{1}{2}\left(\frac{\partial^2}{\partial u^2}+\frac{\partial^2}{\partial v^2}\right)+\frac{1}{2}\left(u^2+v^2\right)+\tilde{g}{\left|\tilde{E}(u,v)\right|}^2\right]\tilde{E}(u,v)\nonumber\\
\nonumber\\
&&=\Lambda\tilde{E}(u,v) \label{twodimensionalgrosspitaevskiiphotons}\,\, ,
\end{eqnarray}
which is formally equivalent to the time-independent Gross-Pitaevskii equation. Here the dimensionless electric field in these units is normalized to unity:
\begin{equation}
\int dudv\,{\left|\tilde{E}(u,v)\right|}^2=1 \,\, .
\end{equation}

\subsection{The Thomas-Fermi limit}
We first review the limit of very dense photons cloud therefore, by analogy with the same limit in atomic clouds, the kinetic term can be neglected, and equation (\ref{twodimensionalgrosspitaevskiiphotons}) reduces to an algebraic equation for the transverse part of the electric field, yielding
\begin{equation}
{\left|\tilde{E}(u,v)\right|}^2=\frac{1}{\tilde{g}}\left[\Lambda-\frac{1}{2}\left(u^2+v^2\right)\right]\theta\left(2\Lambda-u^2-v^2\right) \label{grintfelectricmode}\,\, ,
\end{equation}
where here $\theta(x)$ is the unit step function.
\begin{figure}[t]
\centering
\includegraphics[width=8cm]{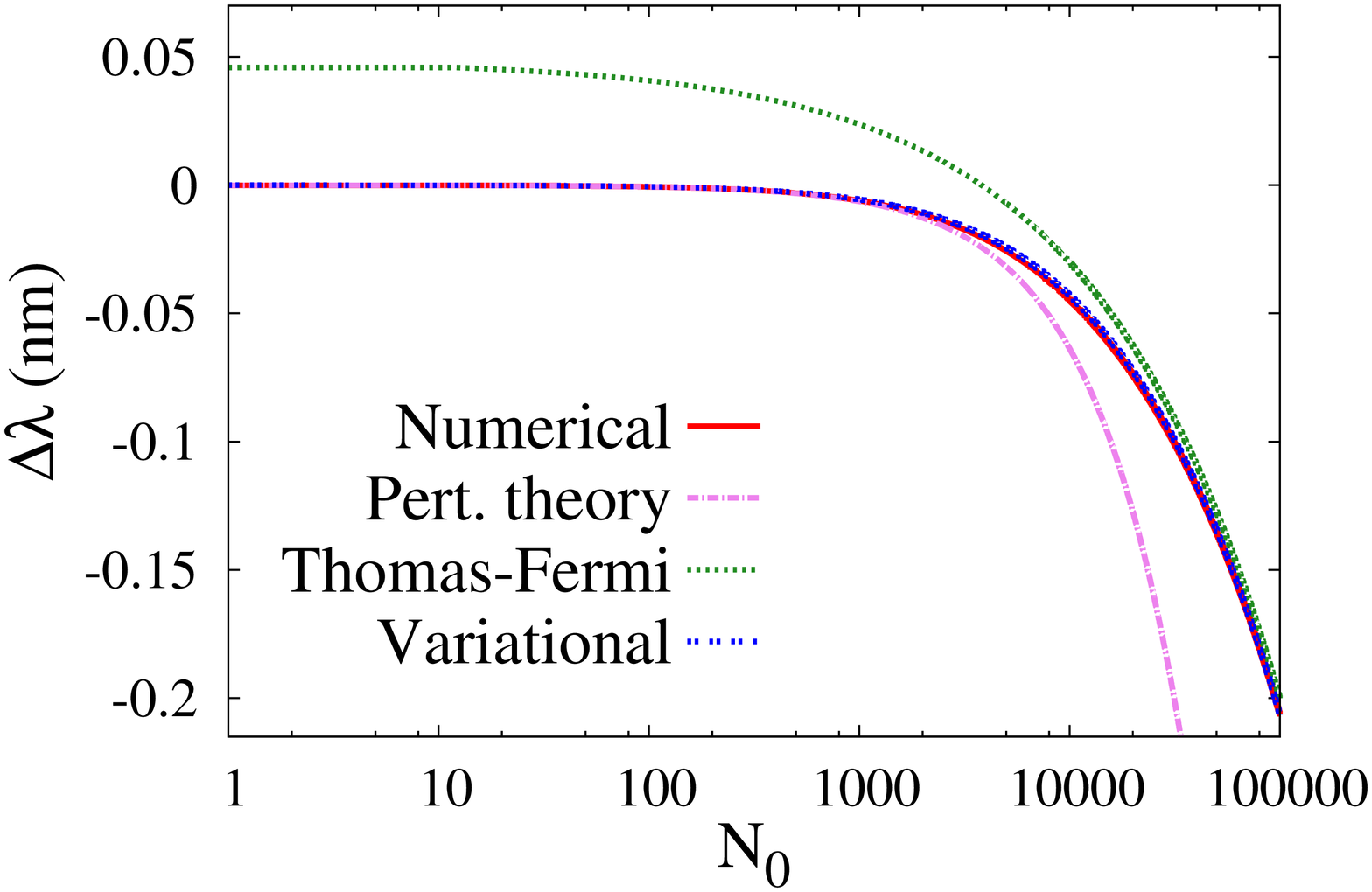}
\caption{(Color online) Shift of the condensate photon wavelength as a function of the number of photons in the condensate obtained by a numerical solution of the local 2D GPE (red full line). The shift in the Thomas-Fermi limit (green dotted line), obtained from (\ref{nonlineargrinfrequency}), the shift for small nonlinearity obtained by the perturbative approach (violet dash-dotted line) and the shift obtained by a variational calculation (blue dash-dotted line) are compared to the numerical solution ($\tilde{g}_N=7.5\times10^{-4}$).}
\label{fig:nonlineargrinthomasfermifrequency}
\end{figure}
The boundary is defined by the condition $u^2+v^2=2\Lambda$. The normalization condition of the dimensionless electric field gives
\begin{equation}
\frac{1}{\tilde{g}}\int dudv\,\left[\Lambda-\frac{1}{2}\left(u^2+v^2\right)\right]\theta\left(2\Lambda-u^2-v^2\right)=1 \,\, ,
\end{equation}
and turning to polar coordinates the integral can be carried out directly; it yields, for a \emph{negative} value of the nonlinear index of refraction (\emph{repulsive} interaction) the only valid solution
\begin{equation}
\Lambda(\mathcal{E})=\sqrt{\frac{\tilde{g}(\mathcal{E})}{\pi}}=k_0{\left(\frac{n_0|n_{2,NL}|}{\varepsilon\pi D}\right)}^{1/2}\sqrt{\mathcal{E}} \label{lambdaandgrelation}\,\, ,
\end{equation}
since $\Lambda$ must be always non-negative; since $\Lambda$ is given by the definition (\ref{helmholtzgrinvalues}) one can obtain an algebraic expression for $k_0$ from which one can obtain the expression for the frequency
\begin{equation}
\nu_q(\mathcal{E})=q\nu_{FSR}{\left[n^2_0-2n_0{\left(\frac{n_{2,L}|n_{2,NL}|}{\varepsilon\pi D}\right)}^{1/2}\sqrt{\mathcal{E}}\right]}^{-1/2} \label{nonlineargrinfrequency}\,\, ,
\end{equation}
where the discretization of $\beta$ as in (\ref{longitudinalmodediscretizationb}) was used; the longitudinal mode number $q$ can be fixed once a single longitudinal mode is selected, hence the frequency of the photons in the condensate depends only on the total energy (number of photons) on the ground mode. A variation of the population of the ground mode implies a shift of the condensate photons frequency (\ref{nonlineargrinfrequency}). When the number of photons increases the frequency is blue-shifted; conversely, if the number of photons decreases the frequency is red-shifted.

The variation of the condensate wavelength as a function of the number of photons obtained from (\ref{nonlineargrinfrequency}) is shown in Fig. \ref{fig:nonlineargrinthomasfermifrequency}; for a condensate occupation up to $N_0\sim10^5$ the predicted shift of the condensate photon wavelength is of order $0.1\,\mathrm{nm}$. For a small nonlinarity Eq. (\ref{nonlineargrinfrequency}) becomes
\begin{equation}
\nu_q(\mathcal{E})\simeq\frac{q\nu_{FSR}}{n_0}\left[1+\frac{1}{n_0}{\left(\frac{n_{2,L}|n_{2,NL}|}{\varepsilon\pi D}\right)}^{1/2}\sqrt{\mathcal{E}}\right] \label{thomasfermicondensateapproximate}\,\, ,
\end{equation}
or, equivalently, in terms of the energy
\begin{equation}
\epsilon_q(\mathcal{E})\simeq\frac{mc^2}{n_0^2}\left[1+\frac{1}{n_0}{\left(\frac{n_{2,L}|n_{2,NL}|}{\varepsilon\pi D}\right)}^{1/2}\sqrt{\mathcal{E}}\right] \label{energytfapproximation}\,\, ,
\end{equation}
where Eq. (\ref{effectivegrinmassandfrequency}) was used. In Eq.~(\ref{nonlineargrinvalue}) it is useful to make explicit the dependence on the number of photons of the ground mode by writing
\begin{equation}
\tilde{g}=\tilde{g}_NN_0 \,\, ,
\end{equation}
which is actually the relevant parameter. The Thomas-Fermi limit is a good approximation when $\tilde{g}_NN_0\gg1$.

The FWHM is found to be given by
\begin{equation}
d=2{\left(\frac{|n_{2,NL}|\mathcal{E}}{n_{2,L}\varepsilon\pi D}\right)}^{1/4}=2\sqrt{\frac{\hbar}{m\Omega\sqrt{\pi}}}\,{\left(\tilde{g}_NN_0\right)}^{1/4} \label{thomasfermigrindiameter}\,\, .
\end{equation}
This power law was also discussed in \cite{klaersweitz}.

An estimation of the nonlinear index of refraction can be done \cite{PhysRevLett.99.043903} by assuming that for thermal effects one usually has $n_2^{(I)}\sim(10^{-14}\div10^{-11})\,\mathrm{m^2/W}$, with $\Delta n=n_2^{(I)}I$ and
\begin{equation}
|n_{2,NL}|=\frac{n_2^{(I)}n_0}{Z_0}\simeq(10^{-17}\div10^{-14})\,\mathrm{m^2/V^2} \label{nonlinearindexvalues}\,\, ,
\end{equation}
where $Z_0$ is the vacuum impedance. A condensate with $d\simeq32\,\mathrm{\mu m}$ and $N_0\simeq5\times10^5$ corresponds to [Eq. (\ref{thomasfermigrindiameter})]
$$
|n_{2,NL}|\simeq3.8\times10^{-16}\,\mathrm{m^2/V^2} \,\, ,
$$
which is consistent with the estimation done in (\ref{nonlinearindexvalues}). With the parameters above, the approximated value of the dimensionless coupling constant is
\begin{equation}
\tilde{g}_N=-\frac{m^3c^4n_{2,NL}}{n_0^5\hbar^2\varepsilon D}\simeq8.24\times10^{-4} \label{grincouplingconstantm}\,\, ,
\end{equation}
in agreement with the value found in the experiment \cite{klaersweitz}. Note that with the values here chosen the approximation given by (\ref{energytfapproximation}) is valid also for $N_0\sim10^8$.

It is useful to write the GP equation in the form commonly adopted in BEC introducing a \emph{condensate wave function} $\psi(\mathbf{r})$ which is
\begin{equation}
\psi(\mathbf{r})=\sqrt{\frac{n^2_0\varepsilon D}{2mc^2N_0}}E(\mathbf{r}) \,\, ,
\end{equation}
normalized such that
\begin{equation}
\int d\mathbf{r}\,{\left|\psi(\mathbf{r})\right|}^2=1 \,\, ,
\end{equation}
and letting
\begin{equation}
\tilde{g}_N=-\frac{m^3c^4n_{2,NL}}{\hbar^2n_0^5\varepsilon D} \,\, ,
\end{equation}
as in (\ref{grincouplingconstantm}). We have from Eq. (\ref{twodimensionalgrosspitaevskiiphotons})
\begin{equation}
\left[-\frac{\hbar^2}{2m}\nabla^2_\perp+\frac{1}{2}m\Omega^2\mathbf{r}^2_\perp+\frac{\hbar^2}{m}\tilde{g}_NN_0{\left|\psi(\mathbf{r})\right|}^2\right]\psi(\mathbf{r})=\mu\psi(\mathbf{r}) \,\, ,
\end{equation}
where $\displaystyle{\nabla^2_\perp=\frac{\partial^2}{\partial x^2}+\frac{\partial^2}{\partial y^2}}$ and $\mathbf{r}^2_\perp=x^2+y^2$; the ``chemical potential'' is defined by
\begin{equation}
\mu=\hbar\Omega\Lambda \,\, .
\end{equation}
One might be interested in comparing the quantities here defined with Ref. \cite{klaersweitz}. The optical Kerr effect is described by
\begin{equation}
n(\mathbf{r})=n_0+n_2I(\mathbf{r})=n_0+\frac{1}{2}n_{2,NL}{|E(\mathbf{r})|}^2 \,\, ,
\end{equation}
where $n_{2,NL}=n_0n_2/Z_0$. The dimensionless coupling constant is related to the round trip time $\tau_{rt}=2Dn_0/c$ and
\begin{equation}
I(\mathbf{r})=\frac{2mc^2}{n^2_0\tau_{rt}}N_0{\left|\psi(\mathbf{r})\right|}^2 \qquad \tilde{g}_N=-\frac{2m^3c^4n_2}{\hbar^2n_0^5\tau_{rt}} \,\, .
\end{equation}

\subsection{Small coupling constant}
For small values of the dimensionless coupling constant $\tilde{g}_NN_0\ll1$ time-independent non-degenerate perturbation theory furnishes the effect of the nonlinearity in the condensate photon frequency and ground mode. A related approach can be found in \cite{Kivshar2001225}. In the unperturbed case for the ground mode one has $\Lambda=1$, and the first order correction can be found with
\begin{equation}
\Lambda(\mathcal{E})=1+\tilde{g}V_0 \,\, ,
\end{equation}
where the first order correction is given by
\begin{equation}
V_0=\int dudv\,{\left|\tilde{E}_0(u,v)\right|}^4=\frac{1}{2\pi} \,\, ,
\end{equation}
and the dimensionless ground state is
\begin{equation}
\tilde{E}_0(u,v)=\frac{1}{\sqrt{\pi}}\exp\left(-\frac{u^2+v^2}{2}\right) \,\, ,
\end{equation}
and then one has
\begin{equation}
\Lambda(\mathcal{E})=1+\frac{\tilde{g}}{2\pi} \,\, .
\end{equation}
By using (\ref{helmholtzgrinvalues}) and (\ref{nonlineargrinvalue}) and for small nonlinearity
\begin{equation}
\beta\simeq k_0n_0-\sqrt{\frac{n_{2,L}}{n_0}}+\frac{k^2_0n_{2,NL}\mathcal{E}\sqrt{n_0n_{2,L}}}{2\varepsilon\pi D} \label{betasmallgcase}\,\, .
\end{equation}
Expression (\ref{betasmallgcase}) can be written in terms of the frequencies by using (\ref{longitudinalmodediscretizationb}), and by defining
\begin{equation}
B(\mathcal{E})=\frac{n_{2,NL}}{c\varepsilon D}\sqrt{\frac{n_{2,L}}{n_0}}\,\mathcal{E} \qquad \nu_0=q\frac{\nu_{FSR}}{n_0}+\frac{\Omega}{2\pi} \,\, ,
\end{equation}
being $\Omega$ the effective trapping frequency defined by (\ref{effectivegrinmassandfrequency}), and note that here $\nu_0$ is the ground mode frequency in the unperturbed case. For small nonlinearity one has
\begin{equation}
\nu_q(\mathcal{E})\simeq\nu_0-\frac{1}{2}\nu^2_0B(\mathcal{E}) \label{nuenonlineargrin}\,\, ,
\end{equation}
and this tends to $\nu_0$ in the limit $B\rightarrow0$ (i.e. $\tilde{g}\rightarrow0$). The effect a of small nonlinear term gives a blue shift in the ground mode frequency, being $\tilde{g}>0$. From (\ref{nuenonlineargrin}) the predicted shift is of the order of $10^{-3}\,\mathrm{nm}$. By defining a \emph{nonlinear frequency}
\begin{equation}
\Omega_{NL}(\mathcal{E})=-\pi\nu^2_0B(\mathcal{E})=-\frac{\pi n_{2,NL}\nu^2_0\mathcal{E}}{c\varepsilon D}\sqrt{\frac{n_{2,L}}{n_0}} \,\, ,
\end{equation}
one eventually obtains from (\ref{nuenonlineargrin}) and (\ref{effectivegrinmassandfrequency})
\begin{equation}
\epsilon_q(\mathcal{E})=\frac{mc^2}{n^2_0}+\hbar\Omega+\hbar\Omega
_{NL}(\mathcal{E}) \,\, ,
\end{equation}
and the nonlinear frequency scales linearly with the nonlinear term, causing a shift in the condensate energy.

One can make an estimation of the ground mode electric field by using the first order correction to the unperturbed modes; by using the cylindrical symmetry it is convenient to perform the expansion using the Laguerre-Gauss modes, which are ($l>0$)
\begin{equation}
\tilde{E}_{ml}(\rho,\theta)=\sqrt{\frac{m!}{\pi\,\Gamma\left(m+l+1\right)}}\,\rho^l\,\mathbb{L}^{(l)}_m\left(\rho^2\right)\,e^{-\rho^2/2+il\theta} \,\, ,
\label{normalizedlaguerregaussexpansion}
\end{equation}
where $\Gamma(x)$ is the gamma function. The total ground mode electric field is then
\begin{equation}
\tilde{E}(\rho,\theta)=\tilde{E}_0(\rho,\theta)-\tilde{g}\sum_{m,l}\frac{V_{ml}}{2m+l}\,\tilde{E}_{ml}(\rho,\theta) \,\, ,
\end{equation}
where the summation is performed with the constraint that $m$ and $l$ are not simultaneously equal to zero (i.e. the summation takes into account all the Laguerre-Gauss modes except the zero order). The expansion coefficients are
\begin{equation}
V_{ml}=\int d\rho d\theta\,\rho {\left[\tilde{E}_0(\rho,\theta)\right]}^3\tilde{E}_{ml}(\rho,\theta) \label{grinpertexpantioncoefficients}\,\, ,
\end{equation}
and by performing the integration with respect to $\theta$ we find
\begin{equation}
V_{ml}=\delta_{l0}\,\frac{2}{\pi}\sum_{k=0}^{m}\frac{{(-1)}^k\,m!}{k!\,(m-k)!\,2^{k+2}} \,\, ,
\end{equation}
where we used the explicit expression of Laguerre polynomials \cite{abramowitzstegun}.

The condensate wave function slightly deviates from a Gaussian, in particular, for $\tilde{g}>0$, the ground mode wave function (always normalized to unity) decreases its peak and broadens, as expected from a repulsive interaction. The perturbative regime remains valid provided that $\tilde{g}_NN_0$ is sufficiently smaller than unity; with the estimated value of $\tilde{g}_N$ one could conclude that a perturbative approach is valid up to $N_0\simeq200$, and above this value Eq. (\ref{twodimensionalgrosspitaevskiiphotons}) has to be solved numerically (as shown for the wavelength shift $\Delta\lambda$ in Fig. \ref{fig:nonlineargrinthomasfermifrequency}).
It is interesting to compare the result here obtained with a variational calculation for the ground mode wave function, as recently done in \cite{2014arXiv1403.5503V}; here the authors, including a contact interaction, considered the energy functional
\begin{eqnarray}
E[\psi(\mathbf{r})]&=&\int d\mathbf{r}\left[\frac{\hbar^2}{2m}{\left|\mathbf{\nabla}\psi(\mathbf{r})\right|}^2+\frac{1}{2}m\Omega^2\mathbf{r}^2{\left|\psi(\mathbf{r})\right|}^2\right]\nonumber\\
\nonumber\\
&+&\frac{g}{2}\int d\mathbf{r}\,{\left|\psi(\mathbf{r})\right|}^4 \label{cpsienergyfunctional}\,\, ,
\end{eqnarray}
and used a Gaussian trial wave function $\psi(\mathbf{r})=\exp\left(-\mathbf{r}^2/2a^2\right)/\sqrt{\pi a^2}$, where $\mathbf{r}$ is the two dimensional position and $a$ is the variational parameter; by using this trial function in the energy functional (\ref{cpsienergyfunctional}) one can find
\begin{equation}
E(a)=\frac{\hbar^2N_0}{2ma^2}+\frac{1}{2}m\Omega^2N_0a^2+\frac{gN^2_0}{4\pi a^2} \label{cpsiqenergy}\,\, ,
\end{equation}
and if (\ref{cpsiqenergy}) is minimized with respect to $a$ one obtains
\begin{equation}
a(N_0)=a_{os}{\left(1+\frac{\tilde{g}N_0}{2\pi}\right)}^{1/4} \label{condensatevariationallocalcase}\,\, ,
\end{equation}
where $\tilde{g}=mg/\hbar^2$ and $a_{os}=\sqrt{\hbar/m\Omega}$. For small condensate fraction one obtains
\begin{equation}
\frac{a(N_0)}{a_{os}}\simeq1+\frac{\tilde{g}N_0}{8\pi} \,\, .
\end{equation}
Both the variational method and perturbation theory provide a very good description of the condensate waist and wavelength shift for sufficiently small occupations of the ground state. In Fig. \ref{fig:nonlineargrinthomasfermifrequency} the condensate photon wavelength shift computed by a numerical solution of the local 2D GPE, by perturbation theory and by the variational approach are compared. The variational model, however, does not consider the fact that the shape of the wave functions changes with the nonlinearity; from a Gaussian (unperturbed case) it \emph{gradually} tends to the Thomas-Fermi approximated wave function; the broadening estimated by the perturbative approach here used is then expected to better describe the broadening of the wave function.

The predicted variation of the condensate diameter as a function of the number of photons in the ground mode for the values of the photon mass $m$ and of the trapping frequency $\Omega$ here considered is, however, quite small since it varies from $d\simeq14.0\,\mathrm{\mu m}$ when $N_0\rightarrow 0$ (unperturbed case) to the value $d\simeq14.1\,\mathrm{\mu m}$ when $N_0\simeq200$. This requires very precise measurements of the number of photons on the ground mode, which might be the most challenging issue, since large fluctuations of the condensate fraction were observed \cite{PhysRevE.88.022132}.

\section{Nonlocal nonlinearity}
\label{sec:nonlocalnonliearity}
When the nonlocal dependence of the index of refraction is considered in the wave equation, the two dimensional Gross-Pitaevskii equation reads as
\begin{widetext}
\begin{equation}
\left[-\frac{1}{2}\left(\frac{\partial^2}{\partial u^2}+\frac{\partial^2}{\partial v^2}\right)+\frac{1}{2}\left(u^2+v^2\right)+N_0\int du'dv'\,\mathcal{K}\left(u,u';v,v'\right){\left|\tilde{E}\left(u',v'\right)\right|}^2\right]\tilde{E}(u,v)=\Lambda\tilde{E}(u,v) \label{finitenonlocalgp}\,\, ,
\end{equation}
\end{widetext}
where the dimensionless integral kernel is related to the original one through
\begin{equation}
\mathcal{K}(u,v)=-\frac{mc^2}{n_0}\frac{2k^2_0}{\varepsilon\xi^2}K(u,v) \,\, .
\end{equation}
The local case is indeed obtained by considering
\begin{equation}
K(u,v)=\frac{n_{2,NL}\xi^2}{2D}\,\delta(u)\,\delta(v) \,\, .
\end{equation}
For all practical purposes, one can take the kernel to be a localized function depending parametrically on a \emph{degree of nonlocality} $\sigma$ which tends to the local case in the limit $\sigma\rightarrow0$
\begin{equation}
K(u,v)=\frac{n_{2,NL}\xi^2}{2D}\,F(u,v;\sigma) \,\, ,
\end{equation}
with the request that in the limit $\sigma\rightarrow0$ the local limit is reached, then
\begin{equation}
\lim_{\sigma\rightarrow0}F(u,v;\sigma)=\delta(u)\,\delta(v) \,\, .
\end{equation}
With this assumption, the dimensionless kernel can be recast as
\begin{equation}
\mathcal{K}(u,v)=\tilde{g}_N\,F(u,v;\sigma) \label{integraldefinition}\,\, ,
\end{equation}
where $\tilde{g}_N$ is the coupling constant given by (\ref{grincouplingconstantm}). With the definition (\ref{integraldefinition}) of the integral kernel, the nonlocal two dimensional Gross-Pitaevskii equation (\ref{finitenonlocalgp}) becomes
\begin{widetext}
\begin{equation}
\left[-\frac{1}{2}\left(\frac{\partial^2}{\partial u^2}+\frac{\partial^2}{\partial v^2}\right)+\frac{1}{2}\left(u^2+v^2\right)+\tilde{g}_NN_0\int du'dv'\,F\left(u-u',v-v';\sigma\right){\left|\tilde{E}\left(u',v'\right)\right|}^2\right]\tilde{E}(u,v)=\Lambda\tilde{E}(u,v)\label{finitenonlocal2dgpe} \,\, .
\end{equation}
\end{widetext}
Equation (\ref{finitenonlocal2dgpe}) has been solved numerically for different values of the coupling constant and the degree of nonlocality by assuming a Gaussian kernel
\begin{equation}
F(u,v;\sigma)=\frac{1}{2\pi\sigma^2}\exp\left(-\frac{u^2+v^2}{2\sigma^2}\right)\label{nonlocal2dkernel} \,\, ,
\end{equation}
as detailed below. The temperature profile (\ref{grinnonlocalprofile}) as a function of the rescaled coordinates can be written as
\begin{eqnarray}
&&\Delta T(u,v)=-n_0^3{\left(\frac{\partial n_0}{\partial T}\right)}^{-1}\,\frac{\hbar\Omega}{mc^2}\,\tilde{g}_NN_0\nonumber\\
\nonumber\\
&\times&\int du'dv'\,F\left(u-u',v-v';\sigma\right){\left|\tilde{E}\left(u',v'\right)\right|}^2 \label{nonlocal2dtemperatureprofile}\,\, .
\end{eqnarray}
As done for the local two dimensional Gross-Pitaevskii equation (\ref{twodimensionalgrosspitaevskiiphotons}), one can compare the numerical results with a variational model, 
and with a perturbative one for small nonlinearity, as discussed in the following sections.

\subsection{Highly nonlocal response}
When the kernel is taken to vary over a characteristic length scale much larger than the dimension of the ground mode electric field, the size of ground mode electric field is the smallest length scale of the system and the electric field can be assumed to be a delta-like function centered at $\mathbf{r}=0$.\cite{Snyder1997} This implies that the index of refraction (\ref{nonlocalgrinindex}) becomes
\begin{equation}
n(\mathbf{r})=n_0-\frac{1}{2}n_{2,L}\mathbf{r}^2+K^{(\varepsilon)}(\mathbf{r})\mathcal{E} \,\, ,
\end{equation}
where the factor $\varepsilon/2$ arising from the electric field normalization has been included in the definition of the kernel without any loss of generality. Note that according to (\ref{grinnonlocalprofile}) the kernel in this limit is closely related to the temperature profile
\begin{equation}
\Delta T(\mathbf{r})\simeq \mathcal{E}{\left(\frac{\partial n_0}{\partial T}\right)}^{-1}K^{(\varepsilon)}(\mathbf{r}) \label{temperatureandkernelrelation}\,\, .
\end{equation}
The temperature profile is locally a parabola, and one can approximate the index of refraction by
\begin{equation}
n(\mathbf{r})\simeq n_0-\frac{1}{2}n_{2,L}\mathbf{r}^2+\left(K^{(\varepsilon)}_0+\frac{1}{2}K^{(\varepsilon)}_2\mathbf{r}^2\right)\mathcal{E} \,\, .
\end{equation}
By introducing the shift of the eigenvalue
\begin{equation}
\delta\Lambda=\frac{k^2_0n^2_0}{2\xi^2}\left[2\frac{K^{(\varepsilon)}_0}{n_0}\mathcal{E}+{\left(\frac{K^{(\varepsilon)}_0}{n_0}\mathcal{E}\right)}^2\right] \,\, ,
\end{equation}
and a dimensionless, non unitary, trapping frequency
\begin{equation}
\tilde{\omega}(\mathcal{E})=\sqrt{1-\frac{K^{(\varepsilon)}_2}{n_{2,L}}\mathcal{E}+\frac{K^{(\varepsilon)}_0}{n_0}\mathcal{E}-\frac{K^{(\varepsilon)}_0K^{(\varepsilon)}_2}{n_0n_{2,L}}\mathcal{E}^2} \,\, ,
\end{equation}
one can rewrite the wave equation in the form
\begin{eqnarray}
&&\left[-\frac{1}{2}\left(\frac{\partial^2}{\partial u^2}+\frac{\partial^2}{\partial v^2}\right)+\frac{\tilde{\omega}^2}{2}\left(u^2+v^2\right)\right]\tilde{E}(u,v)\nonumber\\
\nonumber\\
&&=\left(\Lambda+\delta\Lambda\right)\tilde{E}(u,v) \label{hogrinotherequation}\,\, .
\end{eqnarray}
Equation (\ref{hogrinotherequation}) is actually equation (\ref{hogrinequation}) with the a non unitary frequency and with a shift of the eigenvalue. 
Note that both the dimensionless frequency and the shift of the eigenvalue depend on the nonlinearity through $K^{(\varepsilon)}_0$ and $K^{(\varepsilon)}_2$. 
The condensate photon energy is
\begin{equation}
\epsilon_{q}(\mathcal{E})=\frac{1}{\Delta(\mathcal{E})}\left[\frac{mc^2}{n_0^2}+\hbar\Omega(\mathcal{E})\right] \,\, ,
\end{equation}
where the nonlinear trapping frequency is
\begin{equation}
\Omega(\mathcal{E})=\Omega\,\tilde{\omega}(\mathcal{E}) \label{nonlocalgrintrapping}\,\, ,
\end{equation}
and one defines
\begin{equation}
\Delta(\mathcal{E})=1+\frac{K^{(\varepsilon)}_0}{n_0}\mathcal{E}+\frac{1}{2}{\left(\frac{K^{(\varepsilon)}_0}{n_0}\mathcal{E}\right)}^2 \,\, .
\end{equation}
The ground mode electric field is the Gaussian mode with the waist
\begin{equation}
W(\mathcal{E})=\frac{W}{\sqrt{\tilde{\omega}(\mathcal{E})}} \label{nonlocalgrindimension}\,\, .
\end{equation}
The two quantities $K^{(\varepsilon)}_0$ and $K^{(\varepsilon)}_2$ are related, through (\ref{temperatureandkernelrelation}), to the temperature profile; this can be assumed to be a parabola
\begin{equation}
\Delta T(\mathbf{r})=\Delta T_0\left(1-\frac{\mathbf{r}^2}{R^2}\right) \,\, ,
\end{equation}
where $R$ is the characteristic length scale within which the parabolic approximation is valid, e.g. of order of the condensate dimension; this gives the two expressions
\begin{equation}
K^{(\varepsilon)}_0=\frac{\Delta T_0}{\mathcal{E}}\frac{\partial n_0}{\partial T} \qquad K^{(\varepsilon)}_2=-\frac{2\Delta T_0}{\mathcal{E}R^2}\frac{\partial n_0}{\partial T} \,\, .
\end{equation}
For a defocusing effect, $\partial n_0/\partial T<0$, one has $K^{(\varepsilon)}_0<0$ and $K^{(\varepsilon)}_2>0$, describing then a repulsive interaction between photons. By inspection of (\ref{nonlocalgrintrapping}) and (\ref{nonlocalgrindimension}) one sees that if the nonlinearity is increased, for example by increasing the total energy on the ground mode, the trapping frequency decreases. The condensate diameter, in turn, tends to infinity, which is a consequence of the decrease of the trapping frequency, since for $\Omega=0$ one has the case of free bosons, for which no BEC is expected in two dimensions. The total energy on the ground mode at which the trapping frequency vanishes can be referred to as an \emph{evaporation energy}, for by decreasing the trapping frequency until it vanishes the condensate actually evaporates; this energy is found to be
\begin{equation}
\mathcal{E}_{ev}=\frac{n_{2,L}}{K^{(\varepsilon)}_2} \label{evaporationenergy}\,\, .
\end{equation}
However, this limit is non-physical since this model holds provided that the dimension of the condensate electric field is much smaller than the characteristic spatial extension of the kernel, hence we expect this model to be valid for sufficiently small occupation number of the ground mode.

\subsection{Finite general nonlocal case}
We here show two analytic results which can be obtained by time-independent non degenerate perturbation theory and by a variational approach (see also Appendix \ref{section:appendix}).
\subsubsection{Variational method}
In the variational case one assumes a form of the electric field
\begin{equation}
\tilde{E}(u,v)=\frac{1}{\sqrt{\pi a^2}}\exp\left(-\frac{u^2+v^2}{2a^2}\right)\label{nonlocalgpevariational} \,\, ,
\end{equation}
where $a$ is the variational parameter. Let
\begin{equation}
\tilde{\nabla}=\left(\frac{\partial}{\partial u},\frac{\partial}{\partial v}\right) \qquad \tilde{\mathbf{r}}=(u,v) \,\, ,
\end{equation}
then Eq. (\ref{finitenonlocal2dgpe}) is obtained from the functional
\begin{eqnarray}
\mathcal{F}\left[\tilde{E}\right]&=&N_0\int d\tilde{\mathbf{r}}\,\left[\frac{1}{2}{\left|\tilde{\nabla}\tilde{E}\left(\tilde{\mathbf{r}}\right)\right|}^2+\frac{1}{2}\tilde{\mathbf{r}}^2{\left|\tilde{E}\left(\tilde{\mathbf{r}}\right)\right|}^2\right]\nonumber\\
\nonumber\\
&+&\frac{\tilde{g}_N}{2}N_0^2\int d\tilde{\mathbf{r}}d\tilde{\mathbf{r}}'\,{\left|\tilde{E}\left(\tilde{\mathbf{r}}\right)\right|}^2F\left(\tilde{\mathbf{r}}-\tilde{\mathbf{r}}'\right){\left|\tilde{E}\left(\tilde{\mathbf{r}}'\right)\right|}^2  \,\,\,\,\,\,\label{nonlocal2dgpefunctional}
\end{eqnarray}
with the constraint of the normalization of the electric field, added with the inclusion of the Lagrange multiplier $\Lambda$. One can verify that, with the kernel (\ref{nonlocal2dkernel}) and with the electric field (\ref{nonlocalgpevariational}), the functional (\ref{nonlocal2dgpefunctional}) reduces to a function of the variational parameter
\begin{equation}
\mathcal{F}(a)=N_0\left[\frac{1}{2a^2}+\frac{1}{2}a^2+\frac{\tilde{g}_NN_0}{4\pi\left(a^2+\sigma^2\right)}\right] \label{nonlocal2dafunctional}\,\, .
\end{equation}
By requiring that (\ref{nonlocal2dafunctional}) is minimum with respect to $a$ one can see that the variational parameter, for a given value of $N_0$ and $\sigma$, has to be the solution to the equation
\begin{equation}
a^4(N_0)-1=\frac{\tilde{g}_NN_0\,a^4(N_0)}{2\pi{\left[a^2(N_0)+\sigma^2\right]}^2}\label{nonlocal2dawaistvariational} \,\, ,
\end{equation}
which correctly reduces to (\ref{condensatevariationallocalcase}) for $\sigma\rightarrow0$. Once the behavior of the variational parameter is found, the eigenvalue $\Lambda$ can be found with
\begin{equation}
\Lambda=\int d\tilde{\mathbf{r}}\,\tilde{E}\left(\tilde{\mathbf{r}}\right)\mathcal{H}\left[\tilde{E}\right]\tilde{E}\left(\tilde{\mathbf{r}}\right) \,\, ,
\end{equation}
being the electric field normalized to unity and
\begin{equation}
\mathcal{H}\left[\tilde{E}\right]=-\frac{1}{2}\tilde{\nabla}^2+\frac{1}{2}\tilde{\mathbf{r}}^2+\tilde{g}_NN_0\int d\tilde{\mathbf{r}}'\,F\left(\tilde{\mathbf{r}}-\tilde{\mathbf{r}}'\right){\left|\tilde{E}\left(\tilde{\mathbf{r}}'\right)\right|}^2 \label{nonlocal2dhamiltonianlikeoperator}\,\, ,
\end{equation}
and then one finds
\begin{equation}
\Lambda(N_0)=\frac{1}{2a^2(N_0)}+\frac{1}{2}a^2(N_0)+\frac{\tilde{g}_NN_0}{2\pi\left[a^2(N_0)+\sigma^2\right]} \label{nonlocal2denergyvariational}\,\, .
\end{equation}

\subsubsection{Perturbation theory}
To obtain a prediction for the behavior of the eigenvalue and the waist in the limit of small nonlinearity one can use time-independent non degenerate perturbation theory. In the present case, the perturbation term is the convolution between the integral kernel and the unperturbed ground mode (the Gaussian mode), which is
\begin{eqnarray}
\mathcal{H}'&=&\int du'dv'\,F\left(u-u',v-v';\sigma\right){\left|\tilde{E}_0\left(u',v'\right)\right|}^2\nonumber\\
\nonumber\\
&=&\frac{1}{2\pi\left(\frac{1}{2}+\sigma^2\right)}\exp\left\{-\frac{u^2+v^2}{2\left(\frac{1}{2}+\sigma^2\right)}\right\} \label{nonlocal2gpeperturbationh}\,\, .
\end{eqnarray}
The correction to the unperturbed eigenvalue is then found
\begin{equation}
\Lambda(N_0)=1+\tilde{g}_NN_0\left\langle\tilde{E}_0\right|\mathcal{H}'\left|\tilde{E}_0\right\rangle=1+\frac{\tilde{g}_NN_0}{2\pi\left(1+\sigma^2\right)} \,\, .
\end{equation}
To obtain the correction to the ground mode electric field one can expand over the Laguerre-Gauss modes using the cylindrical symmetry of the Hamiltonian operator (\ref{nonlocal2dhamiltonianlikeoperator}); one sees that only modes with $l=0$ can be coupled, then
\begin{equation}
\tilde{E}(\rho,\varphi)=\tilde{E}_0(\rho)-\tilde{g}_NN_0\sum_{m=1}^{\infty}\frac{V_{m0}}{2m}\tilde{E}_{m0}(\rho,\varphi) \,\, ,
\end{equation}
where the expansion coefficients are found by using the explicit expression of the Laguerre polynomiaAppendix \ref{section:appendix}ls
\begin{eqnarray}
V_{m0}&=&\left\langle\tilde{E}_{m0}\right|\mathcal{H}'\left|\tilde{E}_0\right\rangle\nonumber\\
\nonumber\\
&=&\frac{1}{\pi\left(\frac{1}{2}+\sigma^2\right)}\sum_{k=0}^{m}\frac{{(-1)}^k\,m!\,2^k}{k!\,(m-k)!}{\left[\frac{\frac{1}{2}+\sigma^2}{2\left(1+\sigma^2\right)}\right]}^{k+1} \label{nonlocal2dexpansioncoefficients}
\end{eqnarray}
being $\mathcal{H}'$ the perturbation Hamiltonian given by (\ref{nonlocal2gpeperturbationh}). Note that in the limit $\sigma\rightarrow0$ the expansion coefficients (\ref{nonlocal2dexpansioncoefficients}) reduce to (\ref{grinpertexpantioncoefficients}) found in the local limit.
\begin{figure}[t]
\centering
\includegraphics[width=8cm]{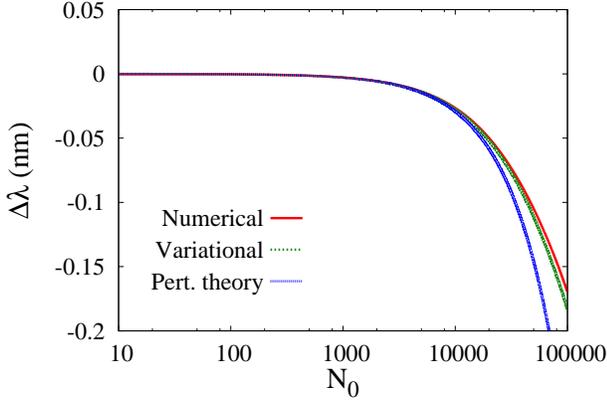}
\caption{(Color online) Condensate photon wavelength shift as a function of the number of photons in the condensate obtained by the numerical simulation of the nonlocal 2D GPE (red full line) compared to the wavelength shift obtained by perturbation theory (blue dotted line) and to the shift obtained by the variational approach (green dashed line, see Appendix \ref{section:appendix}). This figure is obtained with $\tilde{g}_N=7.5\times10^{-4}$ and $\sigma=0.23$.}
\label{fig:condensateshiftnpvh}
\end{figure}
\begin{figure}[t]
\centering
\includegraphics[width=8cm]{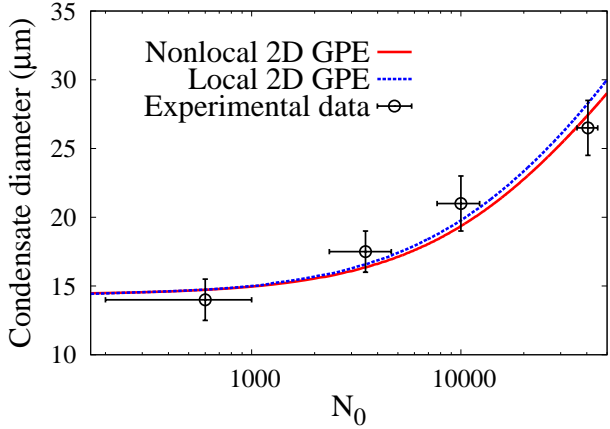}
\caption{(Color online) Experimental points (courtesy of Jan Klaers and Martin Weitz) of the condensate diameter (FWHM) as a function of the number of photons on the ground mode; the data are fitted with the numerical solution of the local 2D GPE (blue dotted line) with $\tilde{g}_N=7.5\times10^{-4}$ and with the numerical solution of the nonlocal 2D GPE with finite nonlocality (red full line) given by Eq. (\ref{finitenonlocal2dgpe}) by considering the kernel as in Eq. (\ref{nonlocal2dkernel}) and with $\tilde{g}_N=(7.5\pm0.1)\times10^{-4}$ and $\sigma=(0.23\pm0.02)$. A quantitative analysis of the two fits shows that the nonlocal solution provides the best fit.}
\label{fig:localandnonlocalfits}
\end{figure}
\begin{figure}[t]
\centering
\includegraphics[width=8cm]{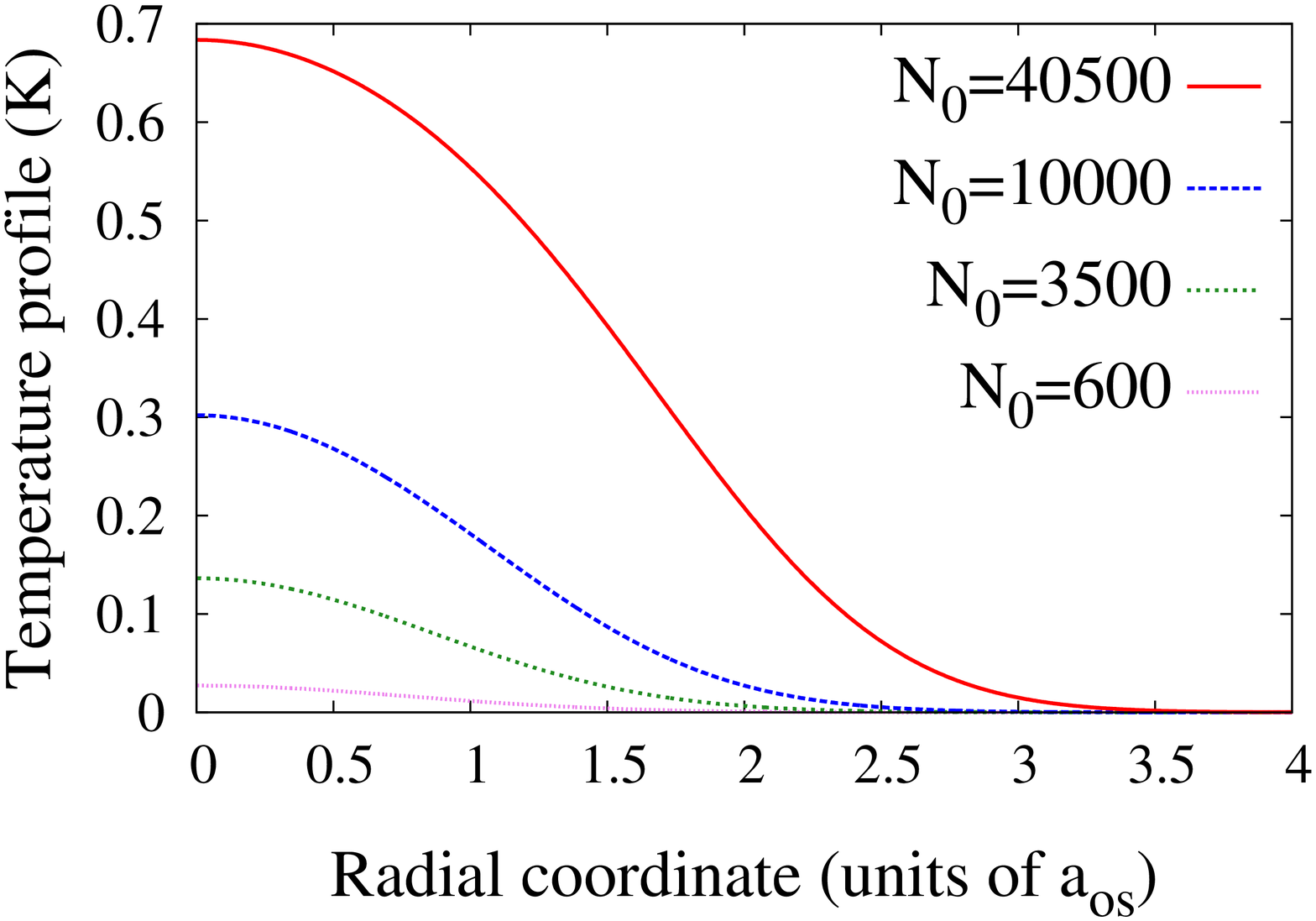}
\caption{(Color online) Radial temperature profile, expressed in Kelvin, as a function of the radial coordinate given in units of $a_{os}=\sqrt{\hbar/m\Omega}\simeq7.8\,\mu m$ with the given parameters, obtained by the numerical solution of the nonlocal 2D GPE by using (\ref{nonlocal2dtemperatureprofile}). In the simulation the kernel was normalized to $\tilde{g}_N$ (with the fit parameters estimated after the best fit in Fig. \ref{fig:localandnonlocalfits}) and the proportionality constant was taken to be $C=n_0^3{\left|\partial n_0/\partial T\right|}^{-1}\hbar\Omega/mc^2\simeq0.22\,K$ since $\Omega=2.6\times10^{11}\,rad/s$, $m=6.7\times10^{-36}\,kg$ and $n_0=1.33$, $\partial n_0/\partial T=-4.86\times10^{-4}\,K^{-1}$ (methanol \cite{LustyDunn:87}) were used in the simulation. The temperature profile is here given for four values of the number of photons, corresponding to the occupancies at which the experimental data fitted in Fig. \ref{fig:localandnonlocalfits} were taken.}
\label{fig:nonlocal2dtemperatureprofile}
\end{figure}
\subsection{Numerical solution and comparison with experiments}
In Fig. \ref{fig:condensateshiftnpvh} we report the wavelength shift $\Delta\lambda$ by numerically solving Eq. (\ref{finitenonlocal2dgpe}) and Eq. (\ref{nonlocal2dkernel}), as detailed in Appendix \ref{section:appendix}. We compare the numerical results with the variational and the perturbative approaches. In Fig. \ref{fig:localandnonlocalfits} we show the FWHM.

One can use the finite nonlocal model to fit the reported experimental data, by using the coupling constant $\tilde{g}_N$ and the degree of nonlocality $\sigma$ as fit parameters. We use the four experimental points in Ref. \cite{klaersweitz}. One can obtain the FWHM $d$ as a function of the number of photons on the ground mode for a given set of $\tilde{g}_N$ and $\sigma$. For a given numerical solution $d\left(N,\tilde{g}_N,\sigma\right)$ the best fit is found by requiring the quantity
\begin{equation}
s=\frac{1}{M}\sum_{i=1}^{M}{\left|d_i\left(\tilde{g}_N,\sigma\right)-y_i\right|}^2 \,\, ,
\end{equation}
is minimum with respect to the fit parameters; here $M=4$ is the number of experimental points, $d_i\left(\tilde{g}_N,\sigma\right)$ is the numerical value of the FWHM for $N_i$ photons and $\left\{y\right\}_i$ are the experimental values of the FWHM. The best fit is found for the following values of the fit parameters
\begin{equation}
\tilde{g}_N=(7.5\pm0.1)\times10^{-4} \qquad \sigma=(0.23\pm0.02) \label{bestfit}\,\, .
\end{equation}
\begin{table}[t]
\centering
\begin{tabular}{|c|c|c|}
\hline
$\mathbf{N_0}$ & $\mathbf{\Delta T}$ \textbf{peak (K)} & $\mathbf{\Delta T}$ \textbf{waist} ($\mathbf{\mu m}$) \\ \hline
600 & 0.03 & 8.39 \\ \hline
3500 & 0.14 & 8.89 \\ \hline
10000 & 0.30 & 9.75 \\ \hline
40500 & 0.68 & 12.2 \\ \hline
50000 & 0.76 & 12.7 \\ \hline
\end{tabular}
\caption{Estimated peak and waist of the temperature profile by the numerical simulation; data from $N_0=600$ to $N_0=40500$ are the ones used in Fig. \ref{fig:localandnonlocalfits} and Fig. \ref{fig:nonlocal2dtemperatureprofile}.}
\label{tab:nonlocaltemperatureparameter}
\end{table}
The determined best fit is reported in Fig. \ref{fig:localandnonlocalfits} where the experimental data (courtesy of Jan Klaers and Martin Weitz) are fitted with a local fit (dotted line) with $\tilde{g}_N=7.5\times10^{-4}$ and with the best fit given by the nonlocal model (full line). With this result one can conclude that the nonlocal model here proposed gives a better description of the data compared to the local model, which implies that nonlocal effects could have been present in the performed experiment.

From the numerical simulation also the temperature profile can be determined by using (\ref{nonlocal2dtemperatureprofile}). The estimated temperature profiles for four different values of the number of photons on the ground mode are shown in Fig. \ref{fig:nonlocal2dtemperatureprofile} and the corresponding estimated values of the peak and the waist are tabulated in Table \ref{tab:nonlocaltemperatureparameter}. For small nonlinearity the wave function slightly deviates from a Gaussian form, then the temperature profile for small nonlinearity is also Gaussian:
\begin{equation}
\Delta T(r)\simeq\frac{C}{2\pi}\left(\frac{\tilde{g}_NN}{\frac{1}{2}+\sigma^2}\right)\exp\left\{-\frac{r^2}{2\left(\frac{1}{2}+\sigma^2\right)}\right\} \,\, ,
\end{equation}
being $r$ the radial coordinate in units of $a_{os}=\sqrt{\hbar/m\Omega}$.

\section{Conclusions}
\label{sec:conclusions}
In this manuscript we have theoretically investigated Bose-Einstein condensation of photons.
With respect to the previous analyses, we considered an equivalent setup given by a dye-doped GRIN lens placed between two planar mirrors. We predict a variation of the condensate wavelength, which was not observed in the experiment so far. This variation, however, is of the order of $0.1\,\mathrm{nm}$ (for large condensate fraction), which could have been actually below the experimental sensitivity, and masked by mechanical instabilities, and thus not observable.

We have analysed the possible role of nonlocal nonlinearities present in the case of thermo-optical effects. We first consider an highly nonlocal response and conclude that this model is valid if the occupation number of the ground mode is such that the condensate diameter remains sufficiently close to the linear value. For larger values of the diameter, a finite degree of nonlocality has to be considered. Indeed, the highly nonlocal model, for the reported experiment, does not provide a proper description of the experimental data.  A finite nonlocal model leads us to conclude that the best description of the experimental data is given by the parameters estimated after Eq. (\ref{bestfit}).

In conclusion, we believe that the analysis here reported indicates a role of nonlocality in the BEC of photons.
A rigorous nonlocal model should account the complete thermal properties of the optical device used, which means that the true form of the Green function in Eq. (\ref{nonlocalgrinindex}) obtained by solving the Cauchy problem (\ref{transportequation}) with proper boundary conditions should be determined. This issue will be considered in future work.

\begin{acknowledgments}
We acknowledge fruitful discussions with Martin Weitz, and Jan Klaers for furnishing the experimental data in Ref. \cite{klaersweitz}. We also acknowledge support from ISCRA-CINECA and Sapienza Ricerca 2013.
\end{acknowledgments}

\appendix
\section{Numerical simulation of the nonlocal 2D GPE}
\label{section:appendix}
\subsection{Convolution in cylindrical coordinates}
To simulate the two dimensional Gross-Pitaevskii equation the convolution between the integral kernel and the wave function has to be written in cylindrical coordinates too. Letting 
\begin{equation}
K(\mathbf{r})=K_0e^{-\mathbf{r}^2/2\sigma^2} \,\, ,
\end{equation}
then the convolution between the kernel and a function $f(\mathbf{r})$ is
\begin{eqnarray}
(K*f)(\mathbf{r})&=&\int d\mathbf{R}\,K\left(\mathbf{r}-\mathbf{R}\right)f\left(\mathbf{R}\right)
\nonumber\\
\nonumber\\
&=&K_0\int d\mathbf{R}\,e^{-{\left|\mathbf{r}-\mathbf{R}\right|}^2/2\sigma^2}f\left(\mathbf{R}\right) \,\, ,
\end{eqnarray}
which gives by setting $(K*f)(\mathbf{r})\equiv G(\mathbf{r})$
\begin{eqnarray*}
G(\mathbf{r})&=&K_0e^{-{\left|\mathbf{r}\right|}^2/2\sigma^2}\int d\mathbf{R}\,e^{-{\left|\mathbf{R}\right|}^2/2\sigma^2}e^{\mathbf{r}\cdot\mathbf{R}/\sigma^2}
f\left(\mathbf{R}\right)\nonumber\\
\nonumber\\
&=&K_0e^{-{\left|\mathbf{r}\right|}^2/2\sigma^2}\int d\mathbf{R}\,e^{-{\left|\mathbf{R}\right|}^2/2\sigma^2}e^{\left|\mathbf{r}\right|\left|\mathbf{R}\right|\cos(\theta)/\sigma^2}
f\left(\mathbf{R}\right)
\end{eqnarray*}
where $\theta$ is the angle between $\mathbf{r}$ and $\mathbf{R}$, and turning to cylindrical coordinates, letting the function $f$ be only a function of the radial coordinate, one finds
\begin{eqnarray}
G(r,\varphi)&=&K_0e^{-r^2/2\sigma^2}\int_{0}^{\infty}dR\,Re^{-R^2/2\sigma^2}f(R)\nonumber\\
\nonumber\\
&\times&\int_{0}^{2\pi}d\Phi\,e^{rR\cos(\Phi-\alpha)/\sigma^2} \,\, ,
\end{eqnarray}
and here $\alpha$ is the angle between $\mathbf{r}$ and the $x$ axis; by rewriting the angular part in terms of the relative angle $\theta=\Phi-\alpha$ and using the fact that the function $e^{rR\cos(\theta)}/\sigma^2$ is a periodic function (with period $2\pi$) the angular integral is
\begin{eqnarray}
\int_{0}^{2\pi}d\theta\,e^{rR\cos(\theta)/\sigma^2}&=&
2\int_{0}^{\pi}d\theta\,\cosh\left[\frac{rR\cos(\theta)}{\sigma^2}\right]\nonumber\\
\nonumber\\
&=&2\pi I_0\left(\frac{rR}{\sigma^2}\right) \,\, ,
\end{eqnarray}
where $I_0(z)$ is the modified Bessel function of order zero, which can be defined by the integral representation
\begin{equation}
I_0(z)=\frac{1}{\pi}\int_{0}^{\pi}d\theta\,e^{\pm z\cos(\theta)}=
\frac{1}{\pi}\int_{0}^{\pi}d\theta\,\cosh\left[z\cos(\theta)\right] \,\, .
\end{equation}
The convolution in cylindrical coordinates for a cylindrically symmetric function can be then written as
$$
G(r)=2\pi K_0e^{-r^2/2\sigma^2}\int_{0}^{\infty}dR\,I_0\left(\frac{rR}{\sigma^2}\right)R\,e^{-R^2/2\sigma^2}f(R)
$$
It can be rewritten in a discrete form; let $h$ be the spatial step, then one finds
\begin{equation}
G_m=2\pi K_0e^{-r^2_m/2\sigma^2}h\sum_nI_{mn}r_ne^{-r^2_n/2\sigma^2}f_n \,\, .
\end{equation}
With the expression of the convolution in cylindrical coordinates the nonlocal Gross-Pitaevskii equation in two dimensions can be simulated as it will be shown in the following section.

\subsection{Nonlocal 2D GPE in cylindrical coordinates}
The Gross-Pitaevskii equation in two dimensions in dimensionless coordinates considering an isotropic harmonic potential is
\begin{eqnarray}
&&\left[-\frac{1}{2}\nabla^2+\frac{1}{2}\mathbf{r}^2+N_0\int d\mathbf{R}\,K(\mathbf{r}-\mathbf{R})\psi^2(\mathbf{R})\right]
\psi(\mathbf{r})
\nonumber\\
&=&\mu\psi(\mathbf{r}) \,\, ,
\end{eqnarray}
and using the expression of the convolution in cylindrical coordinates the 2D GPE in cylindrical coordinates becomes
\begin{widetext}
\begin{equation}
\left[-\frac{1}{2}\left(\frac{d^2}{dr^2}+\frac{1}{r}\frac{d}{dr}\right)+\frac{1}{2}r^2+2\pi N_0K_0\exp\left(-\frac{r^2}{2\sigma^2}\right)\int_{0}^{\infty}dR\,I_0\left(\frac{rR}{\sigma^2}\right)R\,\exp\left(-\frac{R^2}{2\sigma^2}\right)f^2(R)\right]f(r)=\mu f(r) \,\, ,
\end{equation}
\end{widetext}
where the Gaussian kernel $K(\mathbf{r})=K_0\exp\left(-\mathbf{r}^2/2\sigma^2\right)$ was used; note that if the kernel is normalized to the value of $\tilde{g}$, in the limit $\sigma\rightarrow0$ the local two dimensional Gross-Pitaevskii equation is obtained. The nonlocal 2D GPE has to be written in its discretized form, and by introducing a ground wave function $\chi$ and the correction $\varphi$ to the wave function $f=\chi+\varphi$ one can find the equation for the correction $\mathbf{A}\mathbf{\varphi}=\mathbf{b}$, where we define the matrix $\mathbf{A}$ ($\mu\equiv E$)
\begin{eqnarray}
A_{mn}&=&H_{mn}-\delta_{mn}E\nonumber\\
\nonumber\\
&+&4\pi N_0K_0h\,\exp\left(-\frac{r_m^2+r_n^2}{2\sigma^2}\right)
I_{mn}r_n\chi_m\chi_n\nonumber\\
\nonumber\\
&+&2\pi N_0K_0h\,\delta_{mn}\,\exp\left(-\frac{r_n^2}{2\sigma^2}\right)B_n \,\, ,
\end{eqnarray}
the vector $\mathbf{b}$
\begin{eqnarray}
b_m&=&E\chi_{mn}-\sum_nH_{mn}\chi_n\nonumber\\
\nonumber\\
&-&2\pi N_0K_0h\,\exp\left(-\frac{r_m^2}{2\sigma^2}\right)\chi_mB_m \,\, ,
\end{eqnarray}
and finally
\begin{equation}
B_m=\sum_pI_{mp}r_p\,\exp\left(-\frac{r_p^2}{2\sigma^2}\right)\chi^2_p \,\, ,
\end{equation}
where $H$ is the unperturbed Hamiltonian
\begin{equation}
H=-\frac{1}{2}\left(\frac{d^2}{dr^2}+\frac{1}{r}\frac{d}{dr}\right)+\frac{1}{2}r^2 \,\, .
\end{equation}
The correction is found with $\mathbf{\varphi}=\mathbf{A}^{-1}\mathbf{b}$. 
\subsection{Perturbation theory and variational approach in the nonlocal case}
The numerical results can be compared to what can be predicted by time-independent non degenerate perturbation theory (small nonlinearity) and by a variational method; perturbation theory provides the expression for the chemical potential and the wave function. The expression of the chemical potential is found to be given by
\begin{equation}
\mu(N)=1+\frac{gNK_0\sigma^2}{1+\sigma^2} \,\, ,
\end{equation}
and the expression of the wave function is
\begin{equation}
\psi(r,\varphi,N)=\psi_0(r)-gN\sum_{m=1}^{\infty}\frac{V_{m0}}{2m}\psi_m(r,\varphi) \,\, ,
\end{equation}
where the expansion coefficients are given by
\begin{equation}
V_{m0}=\frac{2K_0\sigma^2}{\frac{1}{2}+\sigma^2}\sum_{k=0}^{m}\frac{{(-1)}^k\,m!\,2^k}{k!\,(m-k)!}{\left[\frac{\frac{1}{2}+\sigma^2}{2\left(1+\sigma^2\right)}\right]}^{k+1}
\end{equation}
being $\psi_{m0}$ the Laguerre-Gauss modes with $l=0$. The variational approach gives a good estimation of the waist of the wave function, and uses a variational wave function
\begin{equation}
\psi(\mathbf{r})=\frac{1}{\sqrt{\pi a^2}}e^{-\mathbf{r}^2/2a^2} \,\, ,
\end{equation}
where $a$ is the variational parameter, in the energy functional
\begin{eqnarray*}
E[\psi]&=&N\int d\mathbf{r}\,\left[\frac{1}{2}{\left|\nabla\psi(\mathbf{r})\right|}^2+\frac{1}{2}\mathbf{r}^2{\left|\psi(\mathbf{r})\right|}^2\right]\nonumber\\
\nonumber\\
&+&\frac{g}{2}N^2\int d\mathbf{r}d\mathbf{r}'\,K\left(\mathbf{r}-\mathbf{r}'\,\right){\left|\psi(\mathbf{r})\right|}^2{\left|\psi\left(\mathbf{r}'\,\right)\right|}^2 \,\, ,
\end{eqnarray*}
from which one can obtain the expression of the energy as a function of the variational parameter
\begin{equation}
E(a)=N\left[\frac{1}{2a^2}+\frac{1}{2}a^2+\frac{N}{2}\frac{K_0\sigma^2}{a^2+\sigma^2}\right] \,\, ,
\end{equation}
and if this is minimized with respect to $a$ one sees that the variational parameter, for a given value of $N$ and $\sigma$, has to satisfy the equation
\begin{equation}
a^4-1=\frac{NK_0\sigma^2a^4}{{\left(a^2+\sigma^2\right)}^2} \,\, .
\end{equation}
Note that if the kernel normalization and the nonlocality are not independent (which means that the kernel is normalized to $g$, $K_0=g/2\pi\sigma^2$) the results for $\sigma=0$ are the ones obtained for the local 2D GPE, and for $\sigma\rightarrow\infty$ the results are independent of $N$ since in this limit the kernel tends to zero. 

%

\end{document}